\documentclass[11pt]{article}
\usepackage{moriond-photo,epsfig}
\usepackage{graphicx}
\usepackage{floatflt}

\def\be{\begin{equation}}
\def\ee{\end{equation}}
\def\bea{\begin{eqnarray}}
\def\eea{\end{eqnarray}}

\begin{document}
\vspace*{4cm}
\title{Jet Areas, and What They are Good For~\footnote{In collaboration with
Gavin Salam and Gregory Soyez.
Presented at Moriond QCD, La Thuile, Italy, March 2007. To appear in the Proceedings.
}}

\author{ MATTEO CACCIARI }

\address{LPTHE \\Universit\'e P. et M. Curie - Paris 6, Universit\'e D. Diderot
- Paris 7 and CNRS,
Paris, France}

\maketitle\abstracts{
We introduce the concept of the {\sl area} of a jet, and show how it can be used to
perform the subtraction of even a large amount of diffuse noise from hard jets.}

\section{Introduction}
Jet clustering algorithms, which map the particles observed in the final state
of a high-energy collisions into a smaller number of (usually) well defined
objects -- the jets -- are widely used in the study of the properties of strong
interactions. The jets are usually meant to be good proxies of the original
partons (though the detailed relation is more subtle), and by studying them one
tries to probe the underlying dynamics. The reason for using the jets, rather
than directly the observed hadrons, is that they can be construed as
infrared-safe observables: they are therefore amenable to perturbative QCD
predictions, and their sensitivity to non-perturbative phenomena
(hadronisation, underlying event and pileup effects) can either be kept
under control or corrected for.

In this talk we explore the issue of the susceptibility of jets to
contamination from soft radiation distributed in the form of a roughly uniform
and diffuse background. Physical examples are the pileup originated by multiple
minimum bias collisions in high-luminosity hadron colliders like the LHC, the
many particles produced in a central heavy ion collision and, to a lesser
extent, the underlying event given by perturbative and non-perturbative 
QCD radiation whenever strongly-interacting particles are produced at high
energy. We shall argue that this susceptibility can be quantitatively
characterised in terms of the novel concept of {\sl area of a jet}, which we
shall rigorously introduce. In turn, this will suggest a procedure by means
of which such contamination can be subtracted from the jet momentum, so as to
recover -- to a large extent -- its proxy relation with the parton it
originated from. 

Naively, one can think of the jet area as the surface (in the rapidity-azimuth
plane) over which the particles that have been clustered into a given jet are
distributed. One can also assume that the amount of diffuse
background radiation clustered together with the jet will be 
proportional to this area. One could therefore think of determining somehow the
momentum surface density of this noise, $\rho$, and successively subtract
from the jet momentum a quantity given by $\rho$ times the area of the jet.

Before such a program can be implemented in practice, however, the jet area
needs to be defined more rigorously, and a procedure to extract $\rho$ must be
devised. This is done in \cite{css} and \cite{cs} respectively, where
both aspects are introduced and extensively studied.

\section{Jet Area}
The naive vision of the jet area as the surface covered by the particles that
make up the jet quickly turns out to be fallacious: as the particles are
point-like, this area is zero. Drawing some sort of boundary, like for instance
a convex hull -- the minimal set of particles such that all the others are
contained in the polygon drawn through them  -- is also prone to ambiguities:
different jets may overlap, and a region of space might be arbitrarily assigned
to a jet irrespectively of the properties of the clustering algorithm.

\begin{figure}
  \centering
  \includegraphics[height=5cm, width=0.48\textwidth]{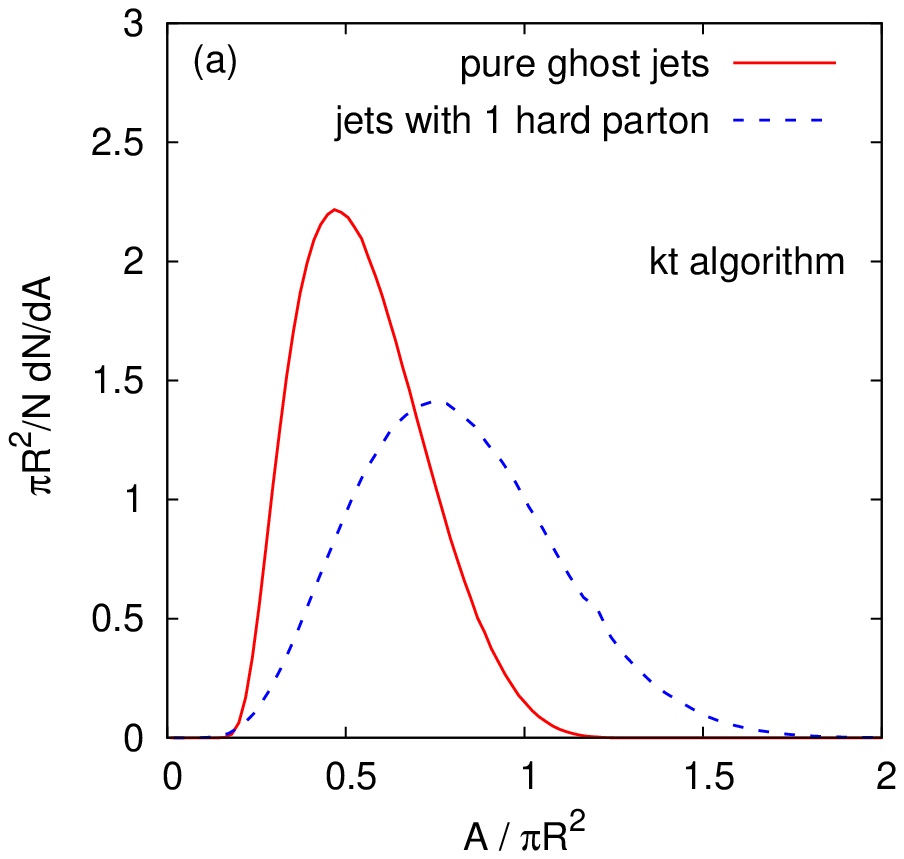}~~~
  \includegraphics[height=5cm, width=0.48\textwidth]{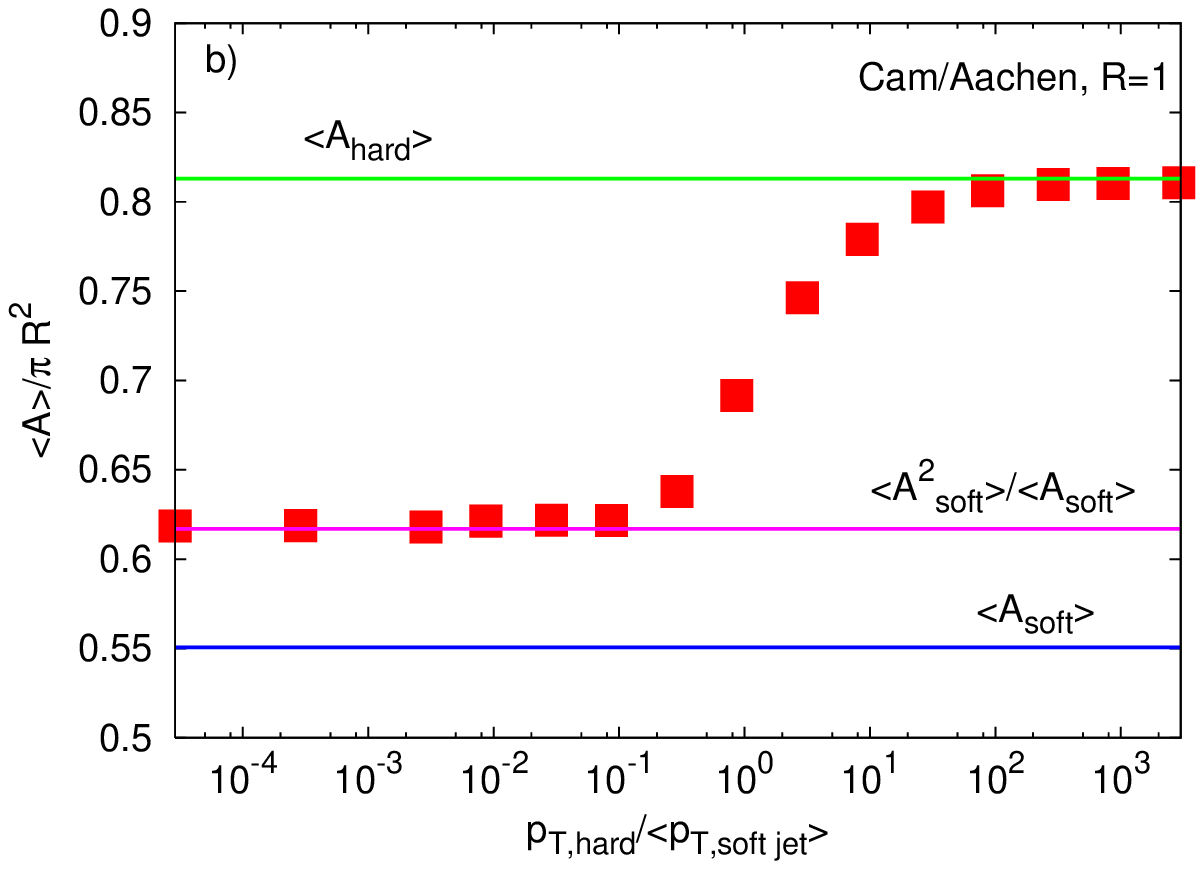}
  \vspace{-.2cm}
  \caption{a) Active area distributions for the $k_t$
  algorithm~\protect\cite{kt}. Cambridge/Aachen~\protect\cite{cam} has a
  very similar behaviour. b) Average are of jet containing a hard
  particle as a function of the ratio of its momentum to that of the soft
  background jets.}
  \label{fig:areas}
 \vspace{-.4cm}
 \end{figure}

To overcome these difficulties, we propose a definition of jet area which is
inherently related to the clustering procedure, and which can properly account
for the jet contamination due to a diffuse background. Our definition is
strictly dependent on the infrared-safety property that a good jet algorithm
should have: the addition of one (or many) soft particles to the event should
not change the final set of hard jets. We add therefore a large number of uniformly
distributed and extremely soft particles ({\sl ghosts}) to the event,
and cluster them together with the real particles. At the end of the clustering
procedure, the number of ghosts clustered with each jet will provide a robust
measure of the jet's extension in the rapidity-azimuth plane, and define
therefore its {\sl active area}, $A$.\footnote{The drawback of this procedure is that a very
large number of particles needs to be clustered (a few thousands ghosts are
needed to achieve accuracies of the order of one per cent). This would be
unfeasible -- or at least extremely unpractical -- without the fast implementations 
of the $k_t$~\cite{kt} and the Cambridge/Aachen~\cite{cam} 
jet algorithms provided by {\tt FastJet}~\cite{Cacciari:2005hq}. 
This package also provides the tools to calculate the area of the jets, as
well as an interface to the new infrared-safe cone algorithm
SISCone~\cite{siscone}.}

Fig.~\ref{fig:areas}(a) shows how the values for this active area are
distributed for two kinds of events: on one extreme, jets constituted
of many uniformly distributed particles with similar momenta (the pure-ghost jets);
on the other extreme, a jet containing a single hard particle. We can
see that these two situations produce different distributions
for the active areas, with different averages: the jets
containing many similar particles have a typical area of
order $\langle A^{soft}\rangle \simeq 0.55\,\pi R^2$ ($R$ is the typical
radius parameter present in most jet algorithms), while the jets
containing a single hard particle tend to be larger, their average area
being  $\langle A^{hard}\rangle \simeq 0.81\, \pi R^2$.

One can take farther this exploration of similarities and differences between
soft (i.e. uniform) and hard jets, and explore how the transition takes place:
fig.~\ref{fig:areas}(b) shows the average area of the jet containing the single
``hard'' particle as its transverse momentum $p_t$ changes from being
negligible with respect to the soft background to being much larger. One can
see that in the $p_{t,\,hard} \gg \langle p_{t,\,soft~jet}\rangle$ limit the
$\simeq 0.81\, \pi R^2$ value for the average area is recovered. On the other
hand, in the opposite  $p_{t,\,hard} \ll \langle p_{t,\,soft~jet}\rangle$ limit
the ``hard'' jet now behaves like a soft one, the difference in average area
being only of probabilistic nature related to the ``measurement'' of the area
of the specific jet containing a given particle.

\begin{figure}[pt]
  \centering
  \includegraphics[height=5cm, width=0.48\textwidth]{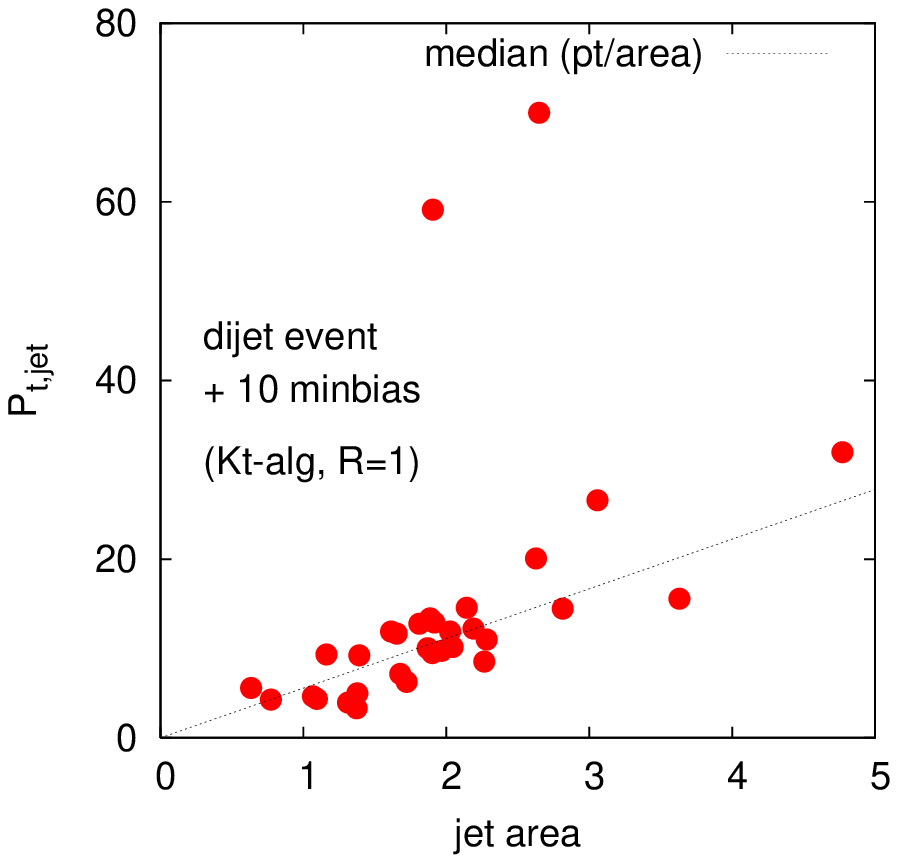}~~
  \includegraphics[height=5cm, width=0.48\textwidth]{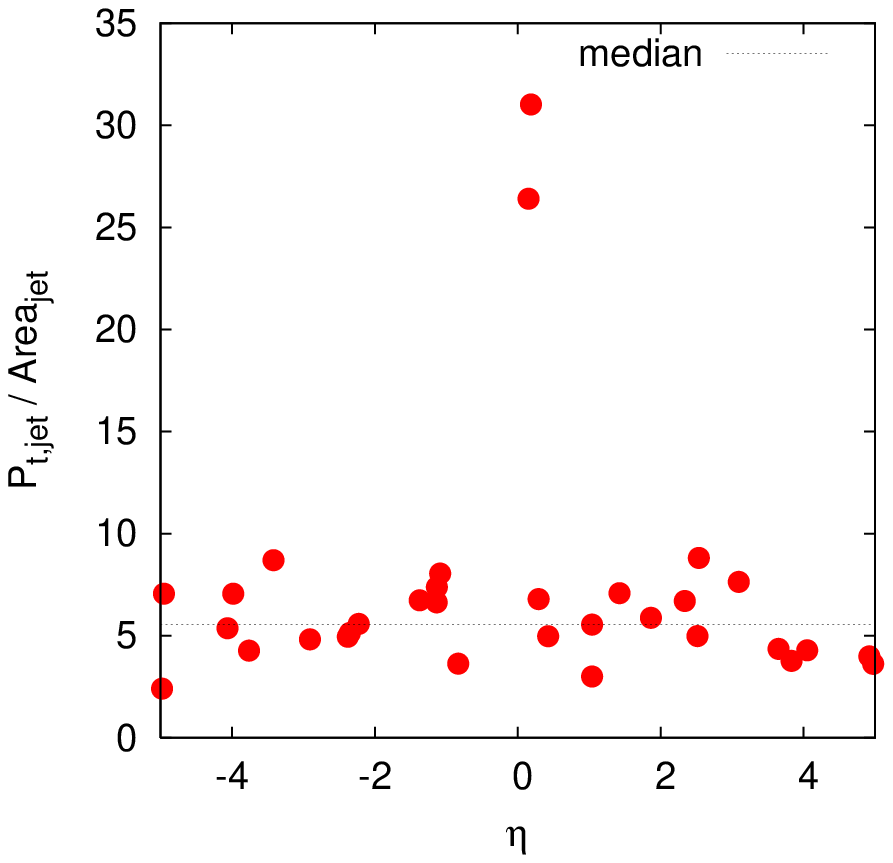}
  \vspace{-.2cm}
  \caption{A dijet event superimposed to 10 minimum bias events originated by
  moderate-luminosity pileup in $pp$ collisions at the LHC, as simulated
  by PYTHIA.}
  \label{fig:rho}
  \vspace{-.4cm}
\end{figure}

\section{Noise Level}
\label{sec:rho}
The estimation of $\rho$, the typical level of the background radiation, could
probably be performed in many ways. The method we propose here is related to
the jet areas discussed above. It relies on the observation that the transverse
momentum of a jet divided by its area, $p_{ti}/A_i$, behaves differently for the
hard jets and for the background ones. Typically, the jets originating from the
background radiation cluster themselves in a band, while the hard jets
stick out. This is clearly shown in fig.~\ref{fig:rho}. This event is a
simulated $pp$ collision at the LHC at moderate luminosity: 10 additional minimum
bias events are added to the main hard collision, which produces a dijet event
with jets of transverse momentum of the order of 50 GeV. Fig.~\ref{fig:rho}
(left) shows that the areas of the various jets can fluctuate widely. However,
when the same jets are plotted in terms of $p_{ti}/A_i$ (right plot) 
one clearly see the
band established by the background. Different strategies can be devised to
quantitatively determine its level. One of the simplest one is to take the
{\sl median} of all the $p_{ti}/A_i$, an operation that prevents
the few hard jets from biasing its value. We define therefore:
\begin{equation}
\label{eq:rho}
\rho = \mathrm{median}\left[\left\{\frac{p_{ti}}{A_i}\right\}\right] \; .
\end{equation}
In the specific case of the event of fig.~\ref{fig:rho}, the momentum density of the
background is therefore $\rho \simeq 6$~GeV per unit area.

\section{Background Subtraction}
Once the area of each jet, $A_i$, and the noise level $\rho$ are known, one can
correct the transverse momentum via the
following operation:
\begin{equation}
\label{eq:subtraction}
p_{ti}^{\mathrm{(sub)}} = p_{ti} - \rho A_i \; .
\end{equation}
We show how this works in practice by considering the following toy model: we
generate many events which contain a single hard particle, with a transverse
momentum $p_t^{hard} = 100$~GeV,  embedded in a background of
10000 soft particles, each with an average transverse momentum $\langle
p_t^{soft}\rangle = 1$~GeV (with little fluctuations, 10\%, around this
value) and
randomly uniformly distributed in rapidity and azimuth up to $y_{max} = 4$. In
this particular case we can of course calculate
the transverse momentum density (per unit area) of the soft  particles
from the input parameters, since we know how we generated them:
\begin{equation}
\label{eq:rho-toy}
\rho = \Big\langle \frac{dp_t^{soft}}{dy\,d\phi} \Big\rangle = \frac {10000
\times 1\,\mathrm{GeV} }{2 \times y_{max} \times 2\pi} \simeq
200~\mathrm{GeV} \, .
\end{equation}
This situation might look extreme, but similar values are expected in realistic
cases, like a central Pb Pb collision at the LHC.

\begin{floatingfigure}{0.48\textwidth}
  \centering
  \includegraphics[height=5cm, width=0.48\textwidth]{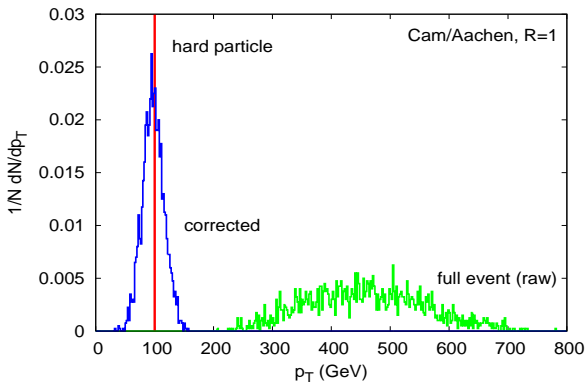}
  \caption{Jets containing a hard particle with $p_{t,\,hard} = 100$~GeV 
  clustered together with a soft background (green, ``raw'' histogram), 
  and after its subtraction (blue ``corrected'' one). The `4-vector'
  versions of the area and of the subtraction~\protect\cite{css,cs}, more
  appropriate for large $R$, have been used
  for this plot.}
  \label{fig:subtraction}
  \vspace{.1cm}
\end{floatingfigure}

We know from the previous section that an average soft jet, when
clustered with the $k_t$ or the Cambridge/Aachen algorithm with $R=1$,
has an area of order $0.55 \pi$. This translates in a typical
transverse momentum $p_t^{soft~jet} \simeq \rho \langle A^{soft}
\rangle \simeq 350~\mathrm{GeV}$. Such jets would already dwarf the
hard particle of 100 GeV. However, this particle will itself be
embedded in a jet containing also many soft particles: this jet will
therefore have a typical transverse momentum of the order of 350 + 100
GeV, but huge fluctuations will be visible from one event to another,
as the amount of background clustered with it will vary considerably. 

This means that both the absolute energy scale and the energy
resolution are degraded by the presence of the background, as shown in
fig.~\ref{fig:subtraction}: the transverse momentum of the hard jet is
displaced, by an amount consistent with our estimate,  and the
resolution is hopelessly bad (green histogram, ``raw'').   However,
once the subtraction is performed according to
eq.~(\ref{eq:subtraction}) (using for each event the $\rho$ directly
extracted from the clustering, as explained in Sec.~\ref{sec:rho}, and
{\sl not} the fixed value of eq.~(\ref{eq:rho-toy}), of course), the
correct average transverse momentum is recovered, together with a large
fraction of the resolution (blue histogram, ``corrected'').

This toy model shows the feasibility and the accuracy of the
determination of the noise level and of the subtraction 
procedure. More realistic examples, and references to experimental
investigations of the problem of background subtraction, are given in \cite{cs}.

{\bf Acknowledgements.} This work has been performed, and is partially in
progress, with Gavin Salam and Gregory Soyez, whom I thank for an extremely
stimulating collaboration.

\end{document}